# Nonparametric Inference of the Population of Compact Binaries from Gravitational-wave Observations Using Binned Gaussian Processes

Anarya Ray[1] , Ignacio Magaña Hernandez[1] , Siddharth Mohite[1] , Jolien Creighton[1] , and Shasvath Kapadia[2]
[1] University of Wisconsin-Milwaukee, Milwaukee, WI 53201, USA; anarya@uwm.edu
[2] The Inter-University Centre for Astronomy and Astrophysics, Post Bag 4, Ganeshkhind, Pune 411007, India; maganah2@uwm.edu

## Abstract

The observation of gravitational waves from multiple compact binary coalescences by the LIGO–Virgo–KAGRA detector networks has enabled us to infer the underlying distribution of compact binaries across a wide range of masses, spins, and redshifts. In light of the new features found in the mass spectrum of binary black holes and the uncertainty regarding binary formation models, nonparametric population inference has become increasingly popular. In this work, we develop a data-driven clustering framework that can identify features in the component mass distribution of compact binaries simultaneously with those in the corresponding redshift distribution, from gravitational-wave data in the presence of significant measurement uncertainties, while making very few assumptions about the functional form of these distributions. Our generalized model is capable of inferring correlations among various population properties, such as the redshift evolution of the shape of the mass distribution itself, in contrast to most existing nonparametric inference schemes. We test our model on simulated data and demonstrate the accuracy with which it can reconstruct the underlying distributions of component masses and redshifts. We also reanalyze public LIGO–Virgo–KAGRA data from events in GWTC-3 using our model and compare our results with those from some alternative parametric and nonparametric population inference approaches. Finally, we investigate the potential presence of correlations between mass and redshift in the population of binary black holes in GWTC-3 (those observed by the LIGO–Virgo–KAGRA detector network in their first three observing runs), without making any assumptions about the specific nature of these correlations.

## 1. Introduction

The first direct observation of gravitational waves (GWs) from the binary black hole (BBH) merger GW150914 (Abbott et al. 2016) by the Laser Interferometric Gravitational Wave Observatory (LIGO; The LIGO Scientific Collaboration et al. 2015) has opened up a new window onto the Universe. Since then, the LIGO–Virgo–KAGRA (LVK; Acernese et al. 2015; The LIGO Scientific Collaboration et al. 2015; Akutsu et al. 2021) detector network has observed about 70 BBHs, with a false alarm rate (FAR) of less than one per year (The LIGO Scientific Collaboration et al. 2021b). Studying the ensemble of BBHs comprised by these detections has facilitated several important investigations, such as the exploration of stellar evolution and binary formation mechanisms (Abbott et al. 2023), the measurement of cosmological parameters (The LIGO Scientific Collaboration et al. 2023a), and tests of General Relativity in the strong-gravity regime (The LIGO Scientific Collaboration et al. 2021c).

In particular, studying the population-level distributions of compact binary coalescences (CBC) masses, spins, and redshifts through the third Gravitational-Wave Transient Catalog (GWTC-3; The LIGO Scientific Collaboration et al. 2021b) has allowed us to empirically probe several models that describe the astrophysical processes responsible for compact binary formation. For example, the existence of a steep falloff in the BBH mass spectrum beyond $50\,M_\odot$ (Fishbach & Holz 2017; Edelman et al. 2021; Abbott et al. 2023) is indicative of the pair-instability process limiting the mass of stellar cores (Fowler & Hoyle 1964; Barkat et al. 1967; Heger & Woosley 2002; Heger et al. 2003; Woosley & Heger 2015; Belczynski et al. 2016; Woosley 2017; Marchant et al. 2019; Renzo et al. 2020). The mass range near which this truncation happens has itself been shown to be informative of nuclear reaction rates in massive stars (Farmer et al. 2020). Similarly, the observed peak in the BBH mass spectrum near the $30\,M_\odot$–$40\,M_\odot$ range (Abbott et al. 2021, 2023, 2023) has been thought to result from the pileup due to pulsational pair-instability supernovae (Woosley 2017; Talbot & Thrane 2018), with the location of the peak expected to be insensitive to stellar metallicity and hence redshift (Farmer et al. 2019). Additional substructure found in the BBH mass spectrum, in the form of peaks and dips atop a smoothed power law (Tiwari & Fairhurst 2021; Edelman et al. 2022a; Edelman et al. 2023; Tiwari 2022), has enabled us to constrain the relative contributions of different formation channels to the BBH population of the Universe. On the other hand, studying the population-level distribution of BBH redshifts through GWTC-3 has led to the discovery that the BBH merger rate increases with redshift (Fishbach et al. 2018, 2021; Karathanasis et al. 2023; van Son et al. 2022a, 2022b; Abbott et al. 2023), shedding light on the metallicity evolution and star formation history of our Universe. Hence, inferring the population properties of CBCs using GW measurements of their system parameters has been highly impactful on our understanding of several astrophysical processes that take place in the Universe.

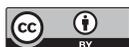

LVK's fourth observing run (O4) can be expected to triple the number of observed BBHs by the end of its first half alone (Abbott et al. 2020a). Hence, analyzing the cumulative catalog of BBH observations post-O4 can enable us to precisely constrain several features in the BBH mass spectrum and redshift distribution. *Parametric* models that assume a specific functional form of the population-level distribution being inferred (Talbot & Thrane 2018; Wysocki et al. 2019; Abbott et al. 2021, 2023; Edelman et al. 2021; Farah et al. 2022; Tiwari 2022) can thus be used to constrain the parameters characterizing these functions with unforeseen precision, given an O4-sized data set, potentially leading to confident empirical validation of the astrophysical assumptions behind such models.

However, a parametric model is inherently restricted, in the sense that it is unable to infer the existence of previously unmodeled features in the underlying population that are beyond its assumptions regarding the functional form of the population distribution. Furthermore, particular features in the functional form assumed by parametric population models can often manifest themselves as a posteriori, even when the data strongly disfavor their existence (Callister et al. 2022; Callister & Farr 2023a), potentially leading to model-induced *false alarms* in the conclusions drawn from parametric population inference. Hence, given the uncertainties regarding the true population distribution and the motivation to search for new physics beyond the assumptions built into existing parametric models, data-driven population inference with minimal suppositions regarding the underlying population is highly important for O4 and beyond.

Several model-independent inference frameworks studied in the existing literature are able to extract features in the CBC population distribution from GW data, without assuming a priori the precise nature and location of these features. Some examples include population modeling based on autoregressive processes (Callister & Farr 2023a), splines (Edelman et al. 2022a; Edelman et al. 2023), Gaussian mixture models (Tiwari 2021, 2022; Tiwari & Fairhurst 2021), adaptive kernel density estimation (aKDE; Sadiq et al. 2022), maximum population likelihood (Payne & Thrane 2023), Dirichlet processes (Rinaldi & Del Pozzo 2021), and binned Gaussian processes (GPs; Foreman-Mackey et al. 2014; Mandel et al. 2016; Mohite 2022; Abbott et al. 2023). While these methods have their individual pros and cons, apart from the aKDE-based one, they all implement a restrictive inference that does not allow for generic *correlations* between the CBC mass and redshift distributions. While most of the other methods can infer the redshift evolution of the combined merger rate either individually or simultaneously with the mass population, they are all based on the simplifying assumption that the shape of the mass spectrum itself does not evolve with redshift. This artifact, when built into a population analysis framework, can prevent it from exploring several astrophysical phenomena, which in fact predict correlations between the shape of the BBH mass spectrum and the distribution of BBH redshifts (Fishbach et al. 2021; van Son et al. 2022a). On the other hand, the existing implementation of the aKDE method has been used to search for mass–redshift correlations only in the detectable population of CBCs, i.e., without accounting for selection biases (Sadiq et al. 2022). Hence, the constraints inferred by Sadiq et al. (2022) cannot be used directly to probe the aforementioned astrophysical phenomena without first converting them into constraints on the astrophysical population (as opposed to the detectable one) by means of appropriately constructed selection functions (Farr 2019; Mandel et al. 2019; Vitale et al. 2020). Hence, previous studies on the existence of correlations in the astrophysical BBH population (such as the parameterized approaches of Callister et al. 2021; Fishbach et al. 2021; Biscoveanu et al. 2022 or the mixture-model-based approaches of Wang et al. 2022; Abbott et al. 2023; Godfrey et al. 2023; Li et al. 2023) have mostly been carried out while making strong assumptions regarding the functional form of said correlations, rendering the inferred constraints susceptible to the previously mentioned limitations of parametric population modeling.

In addition, empirically probing the cosmological evolution of several astrophysical processes requires correlated population inference by means of data-driven frameworks that are free of the limitations of parametric population modeling. For example, the evolution of the initial conditions of zero-age main-sequence stars with cosmic time (Kudritzki & Puls 2000; Belczynski et al. 2010; Brott et al. 2011; Fryer et al. 2012; Dominik et al. 2015; Neijssel et al. 2019; Safarzadeh & Farr 2019; Farrell et al. 2021; Kinugawa et al. 2021; Vink et al. 2021), the preference of dynamical BBH formation environments toward different BH mass ranges at different redshifts (El-Badry et al. 2018; Rodriguez & Loeb 2018; Santoliquido et al. 2020; Romero-Shaw et al. 2021; Weatherford et al. 2021), the dependence of delay time between isolated BBH formation and merger on the corresponding BH masses (Li et al. 2018; Samsing 2018; Mapelli et al. 2019), and the variation of the relative contributions of different BBH formation channels with redshift (Rodriguez & Loeb 2018; Rodriguez et al. 2019; Santoliquido et al. 2020; Yang et al. 2020; Zevin et al. 2021) all predict a BBH mass spectrum whose shape correlates strongly with the corresponding redshift distribution. Many of these predictions are obtained through numerical simulations, with the predicted population distribution lacking an obvious well-defined functional form as required for parametric modeling. Hence, in order to fully explore the aforementioned astrophysical phenomena that govern BBH formation and evolution, nonparametric analysis schemes capable of inferring the correlations between the BBH mass and redshift populations from GW observations through O4 and beyond are of high importance and significance.

In this work, we develop a model-independent inference framework based on binned GPs that can infer the population-level distributions of CBC masses and redshifts from GW measurements of these quantities for a sample of BBHs. We allow for correlation between the mass and redshift distributions, while also appropriately accounting for selection biases. We demonstrate the accuracy with which our method can constrain the underlying population by simulating astrophysically motivated fiducial populations of BBHs and realistic measurement uncertainties. We also reanalyze public LVK data for events in GWTC-3 and constrain the BBH mass and redshift distribution without *any* prior assumptions on the shape of these distributions. By particularizing to an uncorrelated model, we show that our method yields constraints that are fully consistent with the fiducial underlying population for simulated sources and those yielded by uncorrelated parametric models for GWTC-3, even though our results show *hints* of new features beyond the standard POWER LAW + PEAK model (Talbot & Thrane 2018). We then use our generalized model to constrain, for the first time, the correlations between the BBH mass and redshift distributions from GWTC-3 in a nonparametric manner.

We note that our work is a *proof of concept* in the sense that we obtain the first nonparametric constraints on the correlations between the population-level distributions of BBH masses and redshifts. Our model thus enables us to probe new physics beyond the scope of existing nonparametric models, which either restrict to an uncorrelated mass–redshift inference (Tiwari 2021, 2022; Tiwari & Fairhurst 2021; Edelman et al. 2022a; Edelman et al. 2023; Callister & Farr 2023a) or attempt to infer those correlations in the detectable population without accounting for selection effects (Sadiq et al. 2022). Hence, straightforward generalizations of our method to simultaneously infer the spin population and other astrophysically significant ensemble properties of CBCs are left as future explorations. Similarly, studying the full CBC mass spectrum using our model through the inclusion of low-mass events in the analyzed data sets is an interesting venture beyond the scope of this paper and is also left for future work.

This paper is organized as follows. In Section 2, we describe the construction of our binned population model, the GP prior, and the hierarchical inference framework used in conjunction to constrain the CBC population from multiple GW observations. In Section 3, we summarize the results obtained by applying our method to real as well as simulated data and discuss their implications. In Section 4, we conclude by summarizing the implications of our method in the context of O4 and beyond, while highlighting potential generalizations, which are left as future projects.

## 2. Methods

In this section, we develop our data-driven clustering algorithm within the framework of Bayesian hierarchical inference to constrain the population-level distributions of CBC parameters from GW observations. We make very few underlying assumptions regarding the functional form of the GW source population. We achieve this by first constructing a binned GP model over the aforementioned population distribution.

### 2.1. Binned Model

In order to cluster across the three-dimensional parameter space of binary component masses and redshift ($m_1, m_2, z$), we first divide the space into $N_b$ bins. We then assume that the merger rate density per unit comoving volume per log-component masses per unit source frame time is constant within a particular bin. Hence the choice of $N_b$ determines the resolution with which our model can distinguish features in the CBC mass–redshift spectrum. For the $\gamma$th bin, the rate density is thus defined as

$$n^\gamma = \frac{dN^\gamma}{d\ln m_1 d\ln m_2 dV_C dt_R}, \quad (1)$$

where $N^\gamma$ is the number of events with masses and redshift lying in the $\gamma$th bin, $V_c$ is the comoving volume, and $t_R$ is the source frame time. If we fix the lower and upper edges of the $\gamma$th bin, denoted by $(l_{m_1}^\gamma, l_{m_2}^\gamma, l_z^\gamma)$ and $(u_{m_1}^\gamma, u_{m_2}^\gamma, u_z^\gamma)$, respectively, then the differential fraction of events with masses $m_1$, $m_2$, and at redshift $z$ can be written in terms of rate densities as

$$\frac{dN}{dm_1 dm_2 dz}(\vec{n}) = \sum_\gamma n^\gamma \frac{1}{m_1 m_2} \frac{dV_c}{dz} \frac{T_{\rm obs}}{1+z} \Theta(m_1 - l_{m_1}^\gamma) \Theta \\ \times (u_{m_1}^\gamma - m_1) \Theta(m_2 - l_{m_2}^\gamma) \Theta(u_{m_2}^\gamma - m_2) \Theta(z - l_z^\gamma) \Theta(u_z^\gamma - z), \quad (2)$$

where $\Theta$ is the Heaviside step function and $T_{\rm obs}$ is the total observation time as measured in the detector frame that relates to the source frame time elapsed during observation $T_R = T_{\rm obs}/(1+z)$. We note that the model in Equation (2) is general in the sense that it can even infer the existence of correlations between the mass and redshift distributions, in contrast to several studies in the existing literature (Edelman et al. 2023; Callister & Farr 2023a) that only allow the total merger rate to vary with redshift. Furthermore, binning up a higher-dimensional parameter space can allow for the straightforward generalization of our model to infer the distribution of other GW parameters, such as spins. However, we note that while the higher-dimensional models are straightforward to formulate, their implementation is susceptible to scalability issues, on which we elaborate further in Section 4.

To summarize, for the binned population model in Equation (2) or any higher-dimensional generalization thereof, constraining the rate densities from GW data amounts to inferring the functional form of the population-level distributions of CBC parameters up to the resolution limit imposed by our choice of binning.

### 2.2. Hierarchical Inference

To constrain the rate densities that characterize our model from multiple GW observations, we employ the framework of Bayesian hierarchical inference (Thrane & Talbot 2019). Bayesian inference of GW data yields measurements of CBC parameters for each event in the form of posterior samples that can be reweighted to the population model of interest. Multiple observations can be treated as independent realizations of an inhomogeneous Poisson process and hence combined hierarchically to yield the likelihood of population-level quantities given the combined data set (Loredo 2004; Mandel et al. 2019; Wysocki et al. 2019). In the context of our binned model, the joint likelihood of rate densities $\vec{n}$ given data $\vec{d}$ from a collection of $N_{\rm obs}$ observations takes the following form:

$$p(\vec{d}|\vec{n}) = e^{-N_{\rm det}(\vec{n})} \prod_i^{N_{\rm obs}} \left\langle \frac{\frac{dN}{dm_1 dm_2 dz}(\vec{n})}{p_{\rm PE}(m_1, m_2, z)} \right\rangle_{{\rm samples},i}, \quad (3)$$

where $\langle \cdot \rangle_{{\rm samples},i}$ denotes an average over posterior samples of ($m_1$, $m_2$, $z$) obtained from the $i$th observation and $N_{\rm det}$ is the number of CBCs expected to be *detectable* as a function of the rate densities. For details regarding the convergence of the Monte Carlo integrals implemented by the aforementioned average over posterior samples, see Appendix A. The implicit assumption in Equation (3) is that every observation in the analyzed set is astrophysical, which requires the imposition of a stringent detection threshold when selecting the list of candidate events to be used in population inference. The existence of this threshold introduces a Malmquist bias in the inferred population, since arbitrary draws of ($m_1, m_2, z$) from Equation (2) are not equally likely to be detectable. Hence, integrating the right-hand side of Equation (2) over ($m_1$, $m_2$, $z$) without accounting for selection effects yields a biased estimate of $N_{\rm det}$.

To account for selection biases, we compute $N_{\rm det}$ by simulating a large fiducial population of CBC signals and injecting them into detector noise realizations. The parameters of the simulated events that pass the detection criteria can be reweighted to our binned population model in order to yield an unbiased estimate of $N_{\rm det}$ (Farr 2019; Mandel et al. 2019; Vitale et al. 2020), and thus we write

$$N_{\rm det}(\vec{n}) = \frac{K_{\rm det}}{K_{\rm draw}} \left\langle \frac{\frac{dN}{dm_1 dm_2 dz}(\vec{n})}{p_{\rm draw}(m_1, m_2, z)} \right\rangle_{\rm samples, det}, \quad (4)$$

where $\langle \cdot \rangle_{\rm samples, det}$ denotes an average over detectable samples of ($m_1$, $m_2$, $z$) and $p_{\rm draw}$ is the fiducial population from which the simulations were generated. The numbers $K_{\rm draw}$ and $K_{\rm det}$ denote the total number of simulations generated and the number of simulated events that pass the detection threshold, respectively. For details regarding the uncertainties in empirically estimating $N_{\rm det}$ from simulations, ways of marginalizing over them, and the corresponding accuracy requirements (Farr 2019), see Appendix B.

The average over samples in Equations (3) and (4) corresponding to each bin is proportional to the rate density of a given bin, with the *constants* of proportionality being precomputable given the relevant samples and a choice of binning. The likelihood in Equation (3) can then be used to infer the rate densities, provided a suitable prior has been imposed on them. For details regarding the calculation of the likelihood and the aforementioned constants, see Appendix A

### 2.3. GP Prior

We choose the prior on logarithmic rate densities to be a stationary GP so as to regularize and smoothen the inferred population distribution in the case of sparse data sets (Foreman-Mackey et al. 2014; Mandel et al. 2016). We represent this prior in the following way:

$$\ln \vec{n} \sim \mathcal{N}(\vec{\mu}, \Sigma), \quad (5)$$

where $\mu$ and $\Sigma$ are the mean and covariance matrix of the GP. For the covariance matrix, we use an exponential quadratic function,

$$\Sigma_{\gamma\gamma'}(\sigma, \lambda) = \sigma^2 \exp\left(-\frac{(c^\gamma - c^{\gamma'})^2}{2\lambda^2}\right), \quad (6)$$

where $\sigma$ controls the amplitude of the covariances, $\lambda$ determines the length scale over which the bins are correlated, and $c^\gamma$ is the ($\log m_1$, $\log m_2$, $z$) coordinate of the $\gamma$th bin center. The quantities ($\vec{\mu}$, $\sigma$, $\lambda$) are treated as *hyperparameters* of the model and are inferred simultaneously with the rate densities.

### 2.4. Hamiltonian Monte Carlo Sampling

The GP prior and the likelihood together yield the joint posterior distribution of the rate densities and the hyperparameters, which takes the following form:

$$p(\vec{n}, \vec{\mu}, \sigma, \lambda | \vec{d}) \propto p(\vec{\mu}, \sigma, \lambda) p(\vec{n} | \vec{\mu}, \sigma, \lambda) p(\vec{d} | \vec{n}), \quad (7)$$

where $p(\vec{n} | \vec{\mu}, \sigma, \lambda)$ is the GP prior on the rate densities and $p(\vec{\mu}, \sigma, \lambda)$ are priors on the hyperparameters (chosen to be broad normal, halfnormal, and lognormal distributions in $\mu$, $\sigma$, and $\lambda$, respectively). The constant of proportionality in Equation (7) is the so-called Bayesian evidence of our binned model in the data and is independent of the rate densities and hyperparameters. The stochastic sampling of the posterior density using Monte Carlo techniques can thus be carried out without computing the evidence, with the posterior samples of the rate densities being sufficient for constructing Bayesian credible intervals of the CBC population distribution.

However, for high-resolution inference, the number of quantities being sampled simultaneously becomes large, leading to an increase in computational cost. In the context of our binned model, the quantities being simultaneously inferred ($\vec{n}$, $\vec{\mu}$, $\sigma$, $\lambda$) span a $D = 2N_b +$ two-dimensional space. Algorithms such as random walk Metropolis or Gibbs sampling, which scale poorly with the dimensionality of the space being sampled (Neal 1993; Homan & Gelman 2014), can potentially render high-resolution population inference computationally prohibitive. For this reason, we sample the posterior in Equation (7) using Hamiltonian Monte Carlo (HMC; Neal 2011; Homan & Gelman 2014), which invokes a computational complexity of $O(D^{5/4})$ per independent sample and is significantly more tractable than the $O(D^2)$ complexity of random walk Metropolis (Creutz 1988) or Gibbs sampling (Homan & Gelman 2014).

We perform HMC sampling by means of the No-U-Turn Sampler (Homan & Gelman 2014) that improves upon standard HMC by efficiently autotuning the step size, as implemented in the `PyMC` software library (Salvatier et al. 2016). Once obtained, the stochastic samples of $\vec{n}$ can be used to reconstruct credible intervals of the differential merger rate density as a function of ($m_1$, $m_2$, $z$), which is expected to contain the underlying population distribution of these CBC parameters with certain posterior probability.

### 2.5. Uncorrelated Inference for Small Data Sets

As mentioned before, the generalized population model in Equation (2) allows for and is able to infer correlations between the mass and redshift distributions of CBCs. However, the simultaneous inference of a large number of quantities from small data sets such as GWTC-3 (or any subpopulation thereof) can be expected to yield uninformative constraints on the marginal distributions. Furthermore, constraints yielded by the fully correlated inference cannot be compared to existing parametric and nonparametric population studies on GWTC-3, all of which restrict to population models comprising uncorrelated mass and redshift distributions. Hence, for analyzing GWTC-3-sized data sets, it is often preferable to restrict our generalized inference by prohibiting correlations between mass and redshift distributions. This is achievable by slightly redefining the binned population model:

$$\frac{dN}{dm_1 dm_2 dz}(\vec{n}_m, \vec{n}_z) = \sum_\alpha n_z^\alpha \frac{dV_c}{dz} \frac{T_{\rm obs}}{1+z} \Theta(z - l_z^\alpha) \Theta(u_z^\alpha - z)$$
$$\times \sum_\beta \frac{n_m^\beta}{m_1 m_2} \Theta(m_1 - l_{m_1}^\beta) \Theta(u_{m_1}^\beta - m_1) \Theta(m_2 - l_{m_2}^\beta) \Theta(u_{m_2}^\beta - m_2),$$
$$(8)$$

where $n_z^\alpha n_m^\beta$ holds the same meaning as the rate density defined in Equation (1). With this correlation-free factoring of the rate densities, the GP prior can also be factored into two independent GPs:

$$\ln \vec{n}_m \sim \mathcal{N}(\vec{\mu}_m, \Sigma_m), \quad (9)$$

$$\ln \vec{n}_z \sim \mathcal{N}(\vec{\mu}_z, \Sigma_z), \quad (10)$$

where both $\Sigma_m$ and $\Sigma_z$ are exponentially quadratic while being conditional on the different hyperparameters: $(\sigma_z, \lambda_z)$ and $(\sigma_m, \lambda_m)$, respectively. This drastically reduces the number of independent quantities being simultaneously inferred, specifically from $N_b = (N_m - 1)^2(N_z - 1)$ to $N_b = (N_m - 1)^2 + N_z - 1$, where $N_m$ and $N_z$ are the numbers of bin edges along the mass and redshift axes, respectively. The smaller number of quantities can be constrained informatively from a GWTC-3-sized data set, with the results eligible for straightforward comparison to existing studies that all carry out uncorrelated parametric and nonparametric population modeling. (We note that the $m_1 \geqslant m_2$ convention that is often adapted in the CBC population inference literature, when implemented in the context of our model, leads to $(N_m - 1)^2$ being replaced by $N_m(N_m - 1)/2$.)

### 2.6. Choice of Binning

Prior to analyzing GW data with our models, we must choose the location and width of the bins along each parameter dimension whose population-level distribution we intend to infer. To select from various possible binning choices, one must weigh resolution against computational cost. A higher number of bins over the same region of parameter space can potentially lead to the identification of new features in the underlying population with increased resolution, while simultaneously increasing the computational cost of the ensuing analyses. In the context of our framework, drawing from a GP prior incurs a computational cost that scales with the total number of bins cubed. For the generalized correlated inference, since the number of bin edges along each parameter dimension contributes multiplicatively to the total number of bins, increasing the number of bins by several factors can lead to intractability, given the current central process unit (CPU)–based implementation of our algorithms. Furthermore, a model with a larger number of bins, when used to analyze the same data set, will converge to the underlying distribution for costlier sampler settings, such as a larger number of samples and walkers, due to an increase in the number of quantities being simultaneously inferred from the same data set. A fully scalable implementation of the described inference framework is part of ongoing development. Once achieved, it will enable a systematic study of the effects of bin choices on the inferred distributions, which is currently limited by the computational cost of the existing implementations.

On the other hand, increasing the number of bins beyond the level of resolution required to identify *all* existing features in the underlying population is not expected to further alter the shape of the inferred distribution significantly. This is because the GP hyperparameters, such as correlation lengths and amplitudes, are simultaneously inferred with the rate densities, from the data themselves. For example, once all the features in the underlying distribution along a particular parameter dimension are extracted, doubling the number of bins along said dimension leaves the inferred posterior of the corresponding length scale unchanged. This is indicative of the GP correlating roughly twice as many bins along that dimension. This results in the consistency of inferred constraints on the underlying population, among the two sets of binning choices. We have verified this for the version of binned GP implemented in Abbott et al. (2023) that infers only the mass distributions of CBCs and is hence computationally much cheaper than the three-dimensional models discussed in this work. We have analyzed simplistic populations of simulated CBCs using the two-dimensional model for two different sets of binning choices, one set having twice as many bins as the other. Both choices of binning yield constraints that are fully consistent with each other and also with the underlying true distribution within uncertainties. These results are discussed in Appendix C.

Given these considerations, we propose the following method of choosing the number of bins. Initial results can be obtained with a preliminary binning choice that satisfies several conditions. For the analyses of GWTC-3 presented in this work, we have chosen bins such that the number of bins in every region of parameter space is at least higher than the number of features in the population that are *expected* in that region, given the findings of existing population studies on the same data set. For example, these expectations can be based on the number of features built into the functional forms of known parametric models that were found to best fit the GWTC-3 data set (Abbott et al. 2023). They can also be based on the findings of existing nonparametric models that were used to carry out uncorrelated population inference on the same data set (Edelman et al. 2023; Callister & Farr 2023a). In our study of GWTC-3, we have taken into account the findings of both parametric and nonparametric population studies to choose our initial set of bins.

Once these initial results are obtained, one can keep on doubling the number of bins along a particular parameter dimension and reanalyzing the same data set, until the length-scale posteriors and the constraints on the rate densities stabilize. However, in the context of the three-dimensional models and their current implementation, even the first iteration of bin refinement can lead to computational intractability. For our proof-of-concept study, we thus focus on the results obtained from the initial binning choice to demonstrate the applicability of our method in producing nonparametric constraints on the mass–redshift correlations in the astrophysical BBH population, the first of their kind. We also focus on validating these constraints with simulation studies, while leaving the higher-resolution iterations by means of a scalable implementation of our framework as part of an upcoming study.

### 3. Results

In this section, we summarize the results obtained upon analyzing real and simulated GW data with our nonparametric model. First, we reanalyze GWTC-3 with the uncorrelated model and compare the resulting constraints with those yielded by astrophysically motivated parametric models. We then validate those results for GWTC-3 by running the uncorrelated model on simulated sources drawn from an uncorrelated fiducial population. We also reanalyze GWTC-3 with the generalized (fully correlated) model and infer nonparametric constraints on the redshift evolution of the shape of the BBH mass spectrum, for the first time. Finally, we test our generalized model on two different sets of simulated sources, one comprising an uncorrelated fiducial population of BBHs, while the other correlated, so as to demonstrate the accuracy of our method in inferring the existence and nature of these correlations between the underlying distribution of BBH masses and redshifts. The code developed to implement these analyses are publicly available in the Python package `gppop`.[3]

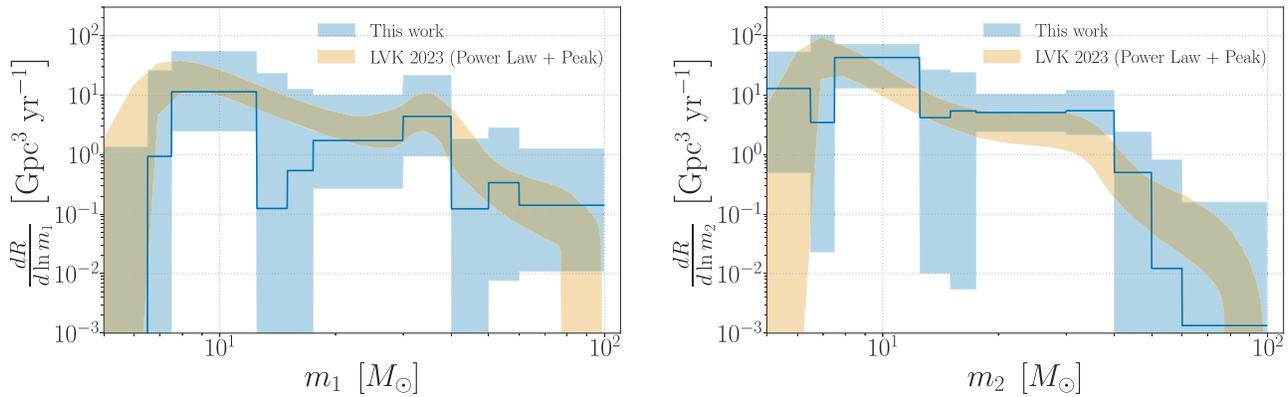

**Figure 1.** Constraints on the CBC mass spectrum from GWTC-3 BBHs using the uncorrelated model. The constraints yielded by our nonparametric model corroborate the findings based on parametric population modeling by Abbott et al. (2023), despite making minimal assumptions about the functional form of the underlying distribution. While there are signs of additional substructure in the primary mass spectrum, for example, near the 15 $M_\odot$ and 55 $M_\odot$ bins, the large error bars on the rate density make any attempt at distinguishing such features from artifacts of Poisson uncertainty inconclusive.

### 3.1. Uncorrelated Inference: GWTC-3

To simultaneously infer the mass spectrum and redshift distribution of BBHs in a model-independent manner, we reanalyze public LVK data (Abbott et al. 2023c) comprising all the BBH events that were observed through GWTC-3 with a FAR of less than one per year. Following previous works, we exclude the *outlier* event GW190814, given the uncertainty regarding its system of origin, which leaves us with a set of 69 high-confidence BBH observations (Abbott et al. 2020b, 2021, 2023; Essick et al. 2022). For each of these events, we use ($m_1$, $m_2$, $z$) parameter estimation (PE) samples calculated directly from the publicly released ($m_1$, $m_2$, $d_L$) samples by LVK to compute Equation (3). Specifically, following Abbott et al. (2023), we convert luminosity distance ($d_L$) samples to redshift by assuming a particular cosmological model, which we choose to be Planck15 (Ade et al. 2016). Since we do not infer the population properties of BBH spins, our inference essentially amounts to fixing the spin population to the default spin distributions used as PE priors. We note that previous implementations of the binned GP model (Abbott et al. 2023) and some semiparametric models (Edelman et al. 2022a) that focused on exploring features only in the mass population of CBCs used the same approach to dealing with spins. To summarize, we use the exact same data set of PE samples used by Abbott et al. (2023), so as to facilitate an apples-to-apples comparison (The LIGO Scientific Collaboration et al. 2021a; The LIGO Scientific Collaboration 2023b). For further details of the single-event PE analyses, see Abbott et al. (2023).

As shown in Figure 1, we find that our uncorrelated model yields constraints on the BBH mass population that are fully consistent with the parametric inference carried out in Abbott et al. (2023) using the POWERLAW+PEAK model, up to measurement uncertainties. It is able to identify both features in the primary mass spectrum, in the form of peaks in the merger rate density at 10 $M_\odot$ and 35 $M_\odot$, which are also found by Abbott et al. (2023)

using the same data set. We also find hints of additional features in the form of a dip near 15 $M_\odot$ and a bump near 65 $M_\odot$. The measurement uncertainty of the inferred population in the bins corresponding to these new features allows for their interpretation as artifacts of Poisson noise. However, given the number of events expected to be observable in O4, the existence of these features can be verified empirically using our nonparametric analysis, unlike parametric models such as POWERLAW+PEAK, which are limited by their assumptions on the functional form of the mass population.

In addition to the mass spectrum, we simultaneously infer the redshift evolution of the BBH merger rate using our uncorrelated model. As shown in Figure 2, our constraints are fully consistent with those obtained using a power law in the $(1 + z)$ model by Abbott et al. (2023) from the same data set, again up to measurement uncertainties. We are able to recover a merger rate that increases with redshift without making any strong assumption regarding the functional form of the evolution. We note that our constraints rule out an unevolving merger rate with less confidence than the parametric model of Abbott et al. (2023), which is to be expected, given the nonparametric nature of our inference. In contrast to the parametric model, our 90% credible intervals are consistent with a nearly unevolving redshift distribution, similar to the findings of other model-independent explorations, such as those based on splines (Edelman et al. 2023) and autoregressive processes (Callister & Farr 2023a). On the other hand, our 68% intervals (dashed lines) are fully inconsistent with a nonincreasing merger rate. Hence, we are able to corroborate that the discovery of an increasing merger rate with redshift is not an artifact of the assumptions built into the parametric model used in that discovery, since we are able to recover the same result, albeit with less confidence. We expect our model to yield more definitive conclusions regarding the redshift evolution of the merger rate given an O4-sized data set, as evident from the results obtained from the simulated catalogs in Section 3.1.1.

#### 3.1.1. Validation of the Uncorrelated Inference

We validate our results obtained from real data using the uncorrelated model by testing our model on mock data sets comprised of simulated sources drawn from a known fiducial population. We choose the underlying distribution of masses to be a truncated power law for both component masses. For the redshift evolution of the merger rate, we choose the underlying distribution to be a power law in $(1+z)$. The intrinsic rate densities for the aforementioned fiducial population thus take the following form:

$$\frac{dR}{dm_1 dm_2}(z) \propto R_0 (1+z)^\kappa m_1^{-\alpha} \frac{\Theta(m_1 - m_2)}{m_1 - m_{\min}}. \quad (11)$$

---

[3] https://github.com/AnaryaRay1/gppop.git

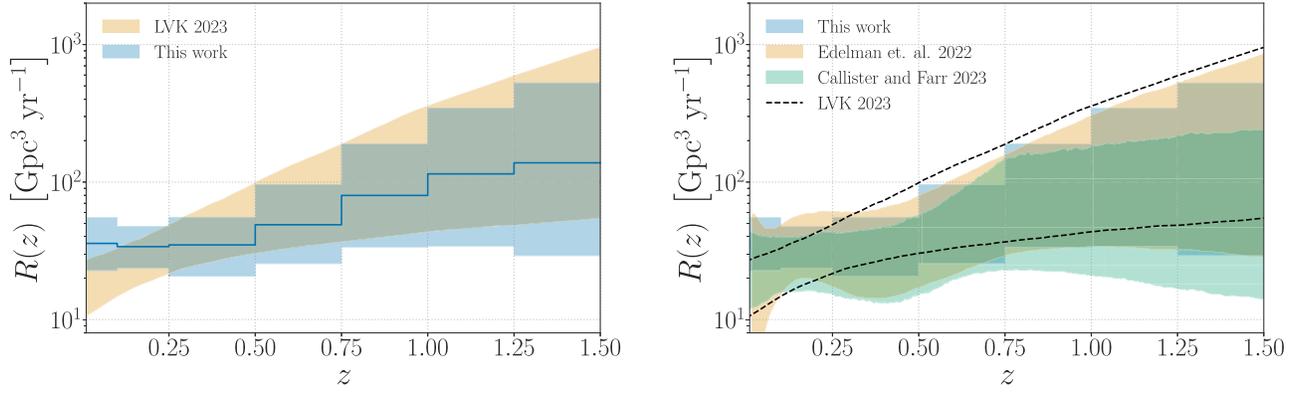

**Figure 2.** Constraints on the redshift evolution of the BBH merger rate from GWTC-3 using the uncorrelated model. Left: comparison of our 90% (shaded) and 68% (dashed) credible intervals with the 90% interval obtained from parametric modeling (Abbott et al. 2023). Right: comparison of our 90% credible intervals with those of other nonparametric studies, such as those based on splines (Edelman et al. 2023, 2022b) and autoregressive processes (Callister & Farr 2023a, 2023b).

The hyperparameter values characterizing our true population are listed in Table 1. After drawing the true values of masses and redshifts for our simulated events from the population using Equation (11), we generate the corresponding *observed* values using realistic estimates of the measurement uncertainties, following the methodology described in Fishbach et al. (2018, 2020) and Farah et al. (2023). We use the advanced LIGO design sensitivity noise curve (Abbott et al. 2020a) to simulate the signal-to-noise ratio (S/N) of mock events as a function of masses and redshifts, in the case where all mock events are assumed to be optimally located and oriented with respect to the detector. We account for the distribution of source orientations and sky positions using a multiplicative factor that encodes information about the antenna response of the detector corresponding to randomly oriented sources, which has been shown to follow a well-modeled distribution for a single detector (Finn & Chernoff 1993). We then interpolate the S/N over a grid of masses and redshifts to generate the PE samples for the simulated events that satisfy a given detection threshold, using the mock PE likelihood described in Farah et al. (2023). We use the same S/N interpolation and detection threshold to generate a different set of detectable simulations required for estimating $N_{\det}$ through Equation (4).

We draw three different realizations of our uncorrelated mock catalog, each spanning a one month observation period with a duty cycle of 0.5, and analyze them individually using the uncorrelated model. With our choice of $R_0$, a single realization is found to comprise a mock catalog of 147 events. The resulting inference, summarized in Figure 3, shows that our model is able to place constraints on the underlying mass population as well as the redshift evolution of the merger rate accurately, up to measurement uncertainty. This validates the results displayed in Figures 1 and 2 obtained from analyzing real data, using the uncorrelated model as being representative of the underlying BBH population as opposed to artifacts of the binned population model itself.

We note that, unlike the case of GWTC-3, the 90% credible intervals on the redshift distribution of simulated events successfully rule out an unevolving merger rate. This is to be expected, given that the simulated catalog has roughly twice as many events as GWTC-3, thereby enabling our model to extract much narrower constraints. Hence, we conclude that our uncorrelated model will be able to confidently constrain the nature of this redshift evolution of the BBH merger rate from an O4-sized catalog.

**Table 1**
True Values for the Hyperparameters Characterizing the Underlying Population in Equation (11)

| Hyperparameter | True Value |
| --- | --- |
| $R_0$ | 100 Gpc$^{-3}$yr$^{-1}$ |
| $\alpha$ | 0.75 |
| $\beta$ | 0.0 |
| $m_{\min}$ | 4.5$M_\odot$ |
| $m_{\max}$ | 55$M_\odot$ |
| $\kappa$ | 3.0 |

### 3.2. Correlated Inference: GWTC-3

To search for correlations between the mass and redshift distributions of BBHs, we reanalyze the GWTC-3 data set as in Section 3.1, but this time with the generalized population model in Equation (2). We constrain the population-level distributions of BBH component masses conditional on redshift, which we obtain from the merger rate density in the following way:

$$p(m_{1,2}|z) = \frac{1}{R(z)} \frac{dR}{dm_{1,2}}(z). \quad (12)$$

We note that uncorrelated population inference frameworks (Edelman et al. 2023; Callister & Farr 2023a) that assume $\frac{dR}{dm_1 dm_2}(z) \propto R_0 f(z) p(m_1, m_2)$ will always recover a distribution $p(m_{1,2}|z)$ that is independent of $z$, regardless of how much the data favor otherwise. On the other hand, our generalized model in Equation (2) has no restrictions built into its underlying assumptions and hence is capable of inferring a potentially evolving $p(m_1|z)$ from the data. We display the inferred credible intervals on $p(m_{1,2}|z)$ at three different redshift bins in Figure 4.

We find that the shapes of the BBH mass spectra at different redshifts are fully consistent with each other up to measurement uncertainty, and hence conclude that there is no evidence of redshift evolution in GWTC-3. However, there are regions where the mass spectrum at lower redshifts has support, while the ones at higher redshifts do not. This is indicative of the fact that we cannot rule out the existence of mass–redshift correlations either, given current observations. On the other hand, given the expectation that the width of the inferred credible intervals scales inversely with the square root of the number of observations, it might be possible to confidently validate or rule out the existence of mass–redshift correlations with our model from an O4-sized data set.

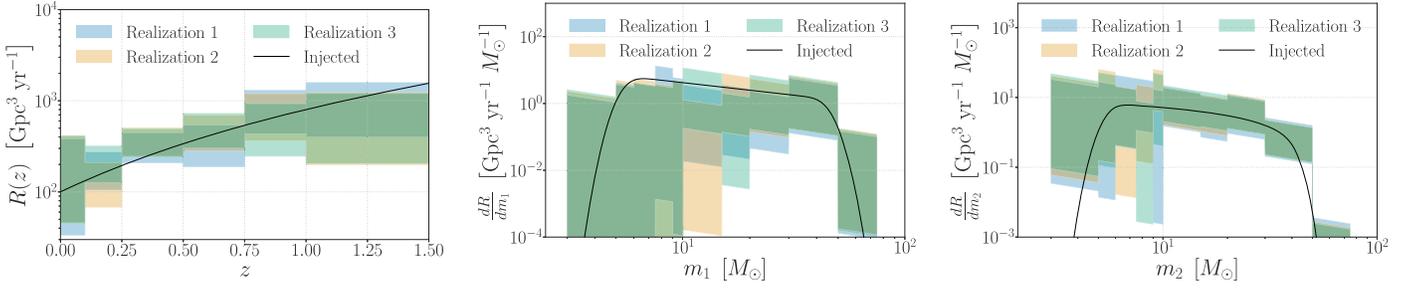

**Figure 3.** Constraints on the underlying population of simulated events using the uncorrelated model. We find that our uncorrelated model is able to recover the true underlying distributions of redshifts and masses up to measurement uncertainty and hence conclude that the results obtained from the real data displayed Figures 1 and 2 are representative of the underlying BBH population and not of any artifacts built into our model construction.

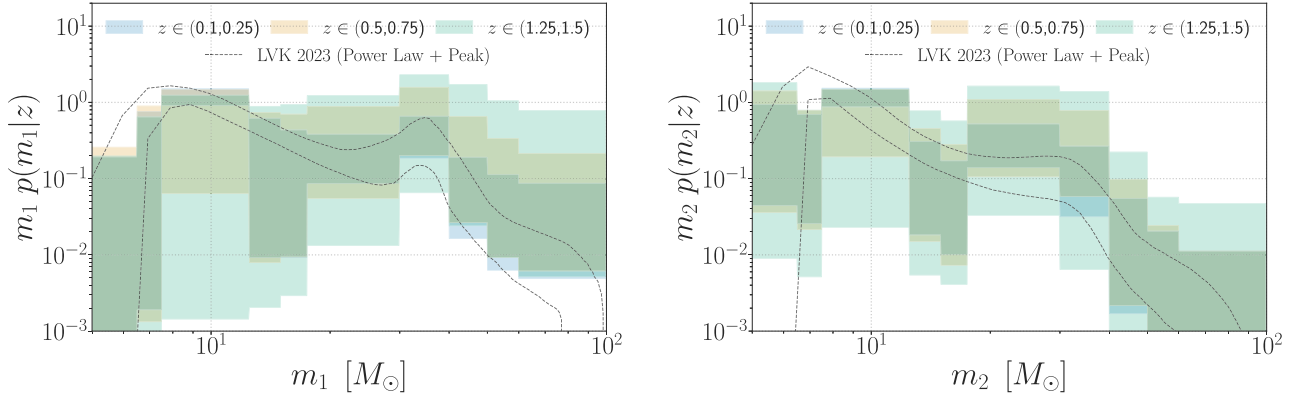

**Figure 4.** Nonparametric constraints on the redshift evolution of the shape of the BBH mass spectrum: the first of their kind. We find that the shapes of the BBH mass distributions across different redshifts agree with each other up to measurement uncertainty. However, there are regions in the BBH component mass ranges where the credible intervals of $p(m_1|z)$ have finite support for some redshift intervals, but not others. Hence, we conclude that it is premature to rule out an evolving BBH mass distribution, given current observations.

*3.2.1. Validation of the Correlated Inference*

We validate our correlated analysis of real data using two different mock data sets, one comprising an uncorrelated mass–redshift population of BBHs and the other a correlated one, so as to demonstrate the accuracy with which our model can infer the nature and existence of mass–redshift correlations in the underlying BBH population. For the first test, we analyze the exact same realizations of the uncorrelated mock observations described in Section 3.1.1, only this time using the fully correlated model in Equation (2). As in the case of real data, we constrain the BBH mass distribution conditional on redshift, so as to demonstrate the accuracy with which our generalized model can recover the underlying densities corresponding to these distributions.

The resulting inference, summarized in Figure 5, demonstrates that our generalized model recovers the shapes of the BBH mass distribution at different redshifts as being fully consistent with the underlying population. In particular, it can be seen that, for all three realizations, the true mass distribution has significant support from the inferred constraints at all redshifts. This validates our model's ability to recover an unevolving mass spectrum from the data without assuming a priori whether or not such evolution may exist. Hence, any potential evidence for (or against) the existence of mass–redshift correlations found using our model from future (post-O4) observations of CBCs can be interpreted as being representative of the underlying CBC population and not as artifacts of the model itself, since the latter would have otherwise manifested in this mock data challenge.

For our second test, we generate a different set of simulated sources that comprise a fiducial population whose mass distribution evolves in shape with redshift. We choose the distribution over the masses to be a power law in primary mass modulated by a Gaussian peak (Talbot & Thrane 2018), with the fraction of events in the Gaussian component varying with redshift. The merger rate density corresponding to the aforementioned underlying population thus takes the following form:

$$\frac{dR}{dm_1 dm_2}(z) \propto R_0 (1+z)^\kappa m_2^\beta \frac{\Theta(m_1 - m_2)}{m_1^{1+\beta} - m_{\min}^{1-\beta}}$$
$$\times \left\{ \frac{m_1^{-\alpha}(1-\alpha)}{m_{\max}^{1-\alpha} - m_{\min}^{1-\alpha}}(1 - \lambda(z)) + \lambda(z)\mathcal{N}_T(m_1, \mu, \sigma, m_{\min}, m_{\max}) \right\},$$
(13)

where $\mathcal{N}_T(m_1, \mu, \sigma, m_{\min}, m_{\max})$ is a truncated Gaussian distribution and $\lambda(z)$ is the fraction of events in the Gaussian component, which is chosen to be a piecewise function of redshift, as in $\lambda(z) = \lambda_0 \Theta(z_0 - z) + \lambda_1 \Theta(z - z_0)$. Here, $\lambda_0$, $\lambda_1$, and $z_0$ are the additional hyperparameters needed to describe a POWERLAW+PEAK mass distribution whose peak fraction evolves with redshift. The chosen fiducial values of all the hyperparameters are listed in Table 2.

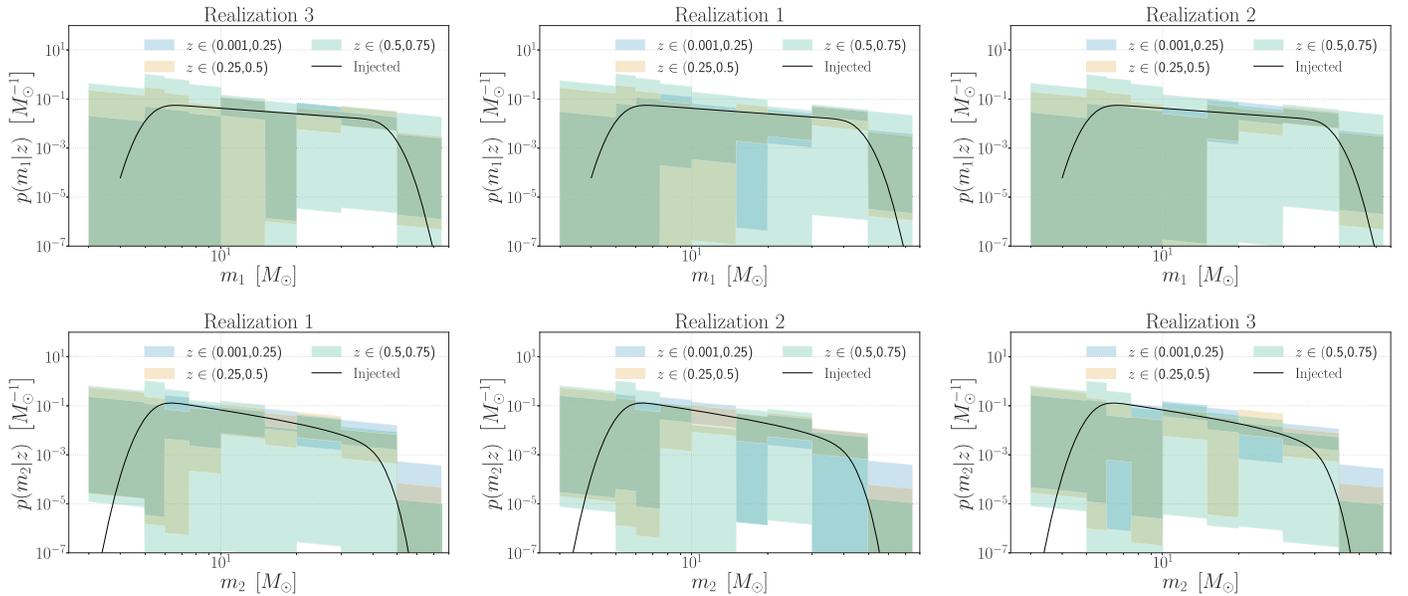

**Figure 5.** Constraints on the mass population of simulated events conditional on redshifts for the mock data set comprising an uncorrelated fiducial population. We find that our correlated model is able to recover an unevolving shape for the underlying distributions of component masses up to measurement uncertainty, and hence conclude that constraints on mass–redshift correlations from the real data displayed in Figure 4 are representative of the true BBH population and not of any artifacts built into our model's construction.

The generation of the mock posterior samples as well as the separate set of detectable simulations for estimating $N_{\rm det}$ is carried out using the exact same methodology and noise curve as for the uncorrelated mock data set, which is described in Section 3.1.1. However, instead of generating different realizations each spanning a small time window, we simulate a much longer observation time, yielding a single realization of 507 mock observations. This choice of simulated run time was made so that the resulting number of mock events above the detection threshold becomes comparable to the upper bound on the expected number of observed BBHs post-O4. This in turn makes our constraints on the mass–redshift correlations inferred from the described mock data set optimistic forecasts for the post-O4 analysis of real data.

Upon analyzing the aforementioned mock data set, we find that our generalized model can correctly recover an underlying BBH mass spectrum that evolves in shape with redshift. As can be seen in Figure 6, our generalized model is able to confidently identify the existence of correlations between the population-level distributions of BBH masses and redshifts, given enough observations. Furthermore, the credible intervals on the BBH mass distribution can be seen to be evolving in complete agreement with the true redshift evolution of the underlying curve. Hence, we conclude that given an O4-sized data set, our nonparametric inference framework can potentially lead to the confident empirical validation of several astrophysical BBH formation models that make informative predictions on the existence of mass–redshift correlations.

### 4. Conclusion and Future Prospects

We have developed a robust and nonparametric hierarchical inference framework based on binned GPs that can constrain the population-level distributions of CBC masses and redshifts from GW data, while allowing for and being able to infer the existence of correlations between the shapes of these distributions. We have demonstrated that our generalized population model has enabled the first nonparametric investigation of the correlations between the underlying distributions of BBH masses and redshifts.

**Table 2**
True Values for the Hyperparameters Characterizing the Underlying Population in Equation (13)

| Hyperparameter | True Value |
|---|---|
| $m_{\rm max}$ | $60 M_\odot$ |
| $m_{\rm min}$ | $6.5 M_\odot$ |
| $\alpha$ | 2.5 |
| $\beta$ | 0 |
| $\mu$ | $35 M_\odot$ |
| $\sigma$ | $4 M_\odot$ |
| $\lambda_0$ | 0.001 |
| $\lambda_1$ | 0.1 |
| $z_0$ | 0.3 |
| $R_0$ | 30 Gpc$^3$ yr$^{-1}$ |

To facilitate comparison with previous works, we have shown that our model can be restricted to one in which the shape of the mass distribution is independent of the redshift evolution of the merger rate. Using the restricted model, we have shown that our method yields measurements of the BBH mass and redshift distributions that are fully consistent with the results of parametric modeling, despite being unassuming of the functional form of these distributions. In addition, using our generalized correlated model, we have inferred for the first time a nonparametric constraint on the redshift evolution of the shape of the BBH mass distribution. We found that even though the credible intervals of the BBH mass distribution at different redshifts are broadly consistent with one another, the large error bars at high redshifts and the very small regions of tension between such intervals make it premature to rule out mass–redshift correlations, given current observations.

We have validated the results from the uncorrelated inference by analyzing a fiducial uncorrelated population of simulated BBHs and realistic measurement uncertainty. In addition, using the same set of simulated sources, we have shown that the correlated inference is capable of correctly recovering an uncorrelated population without assuming anything regarding the existence and shape of these correlations. Last, using a different set of simulated BBHs that comprise a correlated population, we have demonstrated that our generalized model is capable of confidently recovering the correct redshift evolution of the shape of the underlying mass distribution given an O4-sized data set.

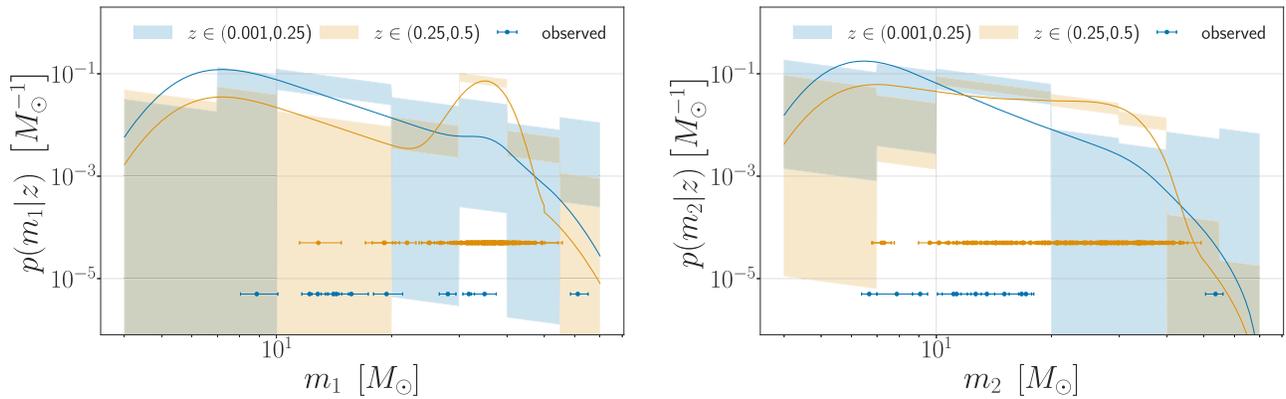

**Figure 6.** Constraints on the mass population of simulated events conditional on redshifts, for the mock data set comprising a correlated fiducial population. We find that our correlated model is able to accurately recover the true underlying distributions of redshifts and masses up to measurement uncertainty. The existence of the disjoint gray and orange regions shows our generalized model can clearly distinguish between the different shapes of the BBH mass spectrum at different redshifts. Hence, we conclude that the results obtained from the real data displayed Figure 4 are representative of the underlying BBH population and not of any artifacts built into our model's construction.

Even though we restrict our work to a mass and redshift population inference, using only BBH observations, we note that such restrictions are straightforward to remove through simple generalizations of our robust and self-consistent inference framework. For example, the ability to simultaneously infer the population-level distribution of BBH spins and its underlying correlations with the corresponding mass and redshift populations can be incorporated within our framework by simply binning up the higher-dimensional space of BBH parameters, now spanned by masses, redshifts, and spins.

However, as mentioned before, repeated draws from the GP algorithm used in this work would incur a computational cost that scales with the total number of bins cubed ($N_b^3$). Hence, adding even a single dimension to the space of the quantities being binned up can potentially lead to a drastic increase in the computational cost of the resulting inference, rendering the current CPU-based implementation intractable. While several workarounds to this issue—such as parallelized computing based on Graphics Process Units, sparse (Quiñ 2005) and scalable (Gardner et al. 2018) GPs, etc.—can potentially be implemented within our framework, the associated developments are beyond the scope of this paper and are hence left as upcoming explorations.

On the other hand, including low-mass events from GWTC-3 in our analysis, thereby fitting across the entire CBC population, is a far simpler endeavor than, say, including spins, amounting to little to no increase in computational cost. However, to facilitate an apples-to-apples comparison between our uncorrelated mass–redshift inference and the corresponding analyses implemented in previous works, such as Edelman & Farr (2023) and Abbott et al. (2023), we have chosen our data set to be exactly identical to the one used by these studies and hence excluded from it the aforementioned collection of low-mass events.

Furthermore, the sparsity of such events compared to BBHs, both in general as well as at most of the higher-redshift bins, is indicative of the fact that such events would have contributed rather uninformatively to the search for population-level correlations between CBC masses and redshifts. While it is indeed of interest and astrophysical significance to study exactly how much the inclusion of these events affects our findings, the corresponding implications and explorations are beyond the scope of this proof-of-concept study and are hence left as follow-up investigations.

To summarize, we have developed a data-driven clustering framework to facilitate robust and self-consistent measurement of the underlying distributions of CBC masses and redshifts and their correlations, from GW observations, for the purpose of exploring new physics beyond the limitations of existing parametric and nonparametric population inference schemes. Armed with our model-independent hierarchical inference algorithm, future studies can hope to empirically probe several highly significant astrophysical phenomena that take place in our Universe in a maximally model-independent manner.

### Acknowledgments

The authors would like to thank Will Farr, Thomas Callister, Amanda Farah, and Soumendra Roy for useful discussions and suggestions. The authors would also like to thank Daniel Wysocki for conducting the LIGO internal review. This work was supported by the National Science Foundation award PHY-2207728. The authors would like to thank an anonymous referee for bringing several references to their attention and for recommending the incorporation of several discussions that have led to the enrichment of the paper. The authors are grateful for computational resources provided by the LIGO Laboratory and supported by the National Science Foundation grants PHY-0757058 and PHY-0823459, PHY-0823459, PHY-1626190, and PHY-1700765. This material is based upon work supported by NSF's LIGO Laboratory, which is a major facility fully funded by the National Science Foundation.

## Appendix A
## The Posterior

In this appendix, we summarize in detail how the various quantities required to evaluate the likelihood in Equation (3) are computed from the GW posterior samples. We also provide an expression for the population posterior as an explicit function of the rate densities and the hyperparameters. For the $i$th GW observation being analyzed, single-event parameter estimation (carried out for generating GW transient catalogs) yields posterior samples of masses and redshifts (converted from luminosity distance samples using a cosmological model) that can be used to compute the *posterior weights* in each bin (Mohite 2022):

$$w_{d_i}^\gamma = \frac{1}{N_{\text{samples},i}} \sum_{(m_{1j},\, m_{2j},\, z_j)\,\sim\, p(m_1, m_2, z|d_i)} \frac{\delta^{\gamma,\alpha(m_{1j},m_{2j},z_j)} \frac{dV_c}{dz}\big|_{z=z_j} \frac{1}{1+z_j}}{p_{\text{PE}}(m_{1j}, m_{2j}, z_j)\, m_{1j} m_{2j}}, \quad (\text{A1})$$

where $\alpha(m_1, m_2, z)$ is the index of the bin inside which $m_1$, $m_2$, $z$ lie, $p_{\text{PE}}(m_{1j}, m_{2j}, z_j)$ is the fiducial prior used during the single-event PE, $N_{\text{samples},i}$ is the number of posterior samples drawn for the $i$th observation, and $\delta^{\gamma,\gamma'}$ is the Kronecker delta function. Explicitly, the fiducial PE priors used throughout this study are uniform in detector frame mass and luminosity distance squared. We downsample single-event PEs that have a larger number of samples, so as to ensure that each event has an exactly identical number of posterior samples prior to weight computation. These weights can be precomputed for each event corresponding to every bin for computational efficiency. Similarly, the set of detectable simulated events generated for computing $N_{\text{det}}$ can be used to compute the sensitive spacetime hypervolume corresponding to each bin in the following way (Mohite 2022):

$$\langle \text{VT} \rangle^\gamma = \sum_{j\,\sim\,(\text{det,draw})} \frac{\delta^{\gamma,\alpha(m_{1j},m_{2j},z_j)} T_{\text{obs}}}{K_{\text{draw}}} \\ \times \frac{\frac{dV_c}{dz}\big|_{z=z_j} \frac{1}{1+z_j} p(\vec{\lambda}_j|\text{draw})}{p(m_{1j}, m_{2j}, z_j, \vec{\lambda}_j|\text{draw})\, m_{1j} m_{2j}}, \quad (\text{A2})$$

where $\vec{\lambda}$ are parameters that characterize a GW waveform in addition to masses and luminosity distance (i.e., the parameters that are not binned up), $p(...|\text{draw})$ is the fiducial distribution from which the simulations are drawn, and $j\sim(\text{det,draw})$ signifies simulated events that pass the detection threshold (Farr 2019). The sensitive hypervolumes, like the posterior weights, can also be precomputed for efficiency.

The posterior weights and the hypervolumes can be used to evaluate the likelihood as a fast-evaluating function of the merger rate densities:

$$p(\vec{d}|\vec{n}) = e^{-\sum_\gamma n^\gamma \langle \text{VT}\rangle^\gamma} \prod_i \sum_\gamma w_{d_i}^\gamma n^\gamma. \quad (\text{A3})$$

To obtain an expression for the posterior as an explicit function of the rate densities and the hyperparameters, we must do so for the GP prior as well. The explicit functional form of the log GP prior for an exponential quadratic kernel looks like the following:

$$\log p(\vec{n}|\vec{\mu}, \sigma, \lambda) = -\frac{1}{2} \log |\Sigma(\sigma, \lambda)| \\ -\frac{1}{2}\sum_{\gamma\gamma'} \{(\log n_\gamma - \mu_\gamma)(\log n_{\gamma'} - \mu_{\gamma'}) \Sigma_{\gamma,\gamma'}^{-1}(\sigma, \lambda)\}, \quad (\text{A4})$$

$$\Sigma_{\gamma\gamma'}(\sigma, \lambda) = \sigma^2 \exp\left[-\sum_{\theta \in \{m_1, m_2, z\}} \frac{(c_{\gamma,\theta} - c_{\gamma',\theta})^2}{2\lambda_\theta^2}\right], \quad (\text{A5})$$

where $|A|$ is the determinant of the matrix $A$. In terms of the likelihood in Equation (A3) and the explicit form of the GP prior, the log of the posterior in Equation (7) can be written as:

$$\log p(\vec{n}, \vec{\mu}, \sigma, \vec{\lambda}|\{w_d\}, \langle \text{VT}\rangle) \\ = \log p(\vec{\mu}, \sigma, \vec{\lambda}) - \frac{1}{2}\log|\Sigma(\sigma, \lambda)| \\ -\frac{1}{2}\sum_{\gamma,\,\gamma'} \{(\log n_\gamma - \mu_\gamma)(\log n_{\gamma'} - \mu_{\gamma'})\Sigma_{\gamma\gamma'}^{-1}(\sigma,\lambda)\} \\ -\sum_\gamma n^\gamma \langle\text{VT}\rangle^\gamma \\ +\sum_i \log\left(\sum_\gamma w_{d_i}^\gamma n^\gamma\right) + \text{const.} \quad (\text{A6})$$

Note that while $\{w_d\}$ replaces the conditional dependence of the posterior on $\vec{d}$, the conditional dependence on $\langle \text{VT}\rangle$ was not explicitly mentioned before. However, since $\langle \text{VT}\rangle$ is computed empirically from a data set of simulated sources, our likelihood and hence posterior were always in fact conditional on $\langle \text{VT}\rangle$. The reason for writing it explicitly at this point of the derivation will become clear in the next appendix, wherein we discuss the Monte Carlo uncertainties in $\langle \text{VT}\rangle$. However, prior to investigating the Monte Carlo uncertainties in the selection function, we first describe the convergence of the event-specific Monte Carlo integrals in Equation (A1).

To check for the convergence of event-specific Monte Carlo integrals, we first compute the *variance* of the posterior weight corresponding to each bin:

$$\text{Var}[w_i^\gamma] = \frac{1}{N_{\text{samples},i}} \sum_{(m_{1j},\, m_{2j},\, z_j)\,\sim\, p(m_1,m_2,z|d_i)} \\ \frac{\delta^{\gamma,\alpha(m_{1j},m_{2j},z_j)}\left(\frac{dV_c}{dz}\big|_{z=z_j}\frac{1}{1+z_j}\right)^2}{p_{\text{PE}}^2(m_{1j}, m_{2j}, z_j)\, m_{1j}^2 m_{2j}^2}. \quad (\text{A7})$$

Using these variances, one can construct metrics for testing the convergence of single-event Monte Carlo integrals, such as the number of effective samples:

$$N_{\text{eff},i} = \frac{\sum_\gamma (n^\gamma)^2 \text{Var}[w_i^\gamma]}{\left(\sum_\gamma n^\gamma w_i^\gamma\right)^2}. \quad (\text{A8})$$

Following Callister & Farr (2023a), to ensure the convergence of the event-specific Monte Carlo integrals, we demand that for each draw of $\vec{n}$ from the hyperposterior, the condition $\min_i \log_{10} N_{\text{eff},i} > 0.6$ be satisfied. However, given the size of the bins chosen for our study and the hyperpriors on the covariance amplitude, we expect these conditions to hold for most draws from the hyperposterior. Hence, instead of penalizing the likelihood upon the violation of said condition, we verify in post-processing that for all rate density samples this condition is automatically satisfied. On the other hand, for future studies that aim to implement inference at a much higher bin resolution, for efficient sampling, we propose that this condition is imposed during sampling in the form of a steep penalty on the likelihood, similar to what is done in Callister & Farr (2023a).

## Appendix B
## Uncertainties in ⟨VT⟩ Estimation

Empirically estimated ⟨VT⟩s are subject to Monte Carlo uncertainties (Farr 2019). According to the central limit theorem, the sum over the samples in Equation (A2) can be interpreted as the realization of a Gaussian distribution centered around the true value of the integral being approximated by the mentioned sum. Since we only have a finite number of samples, the Gaussian can be expected to have a finite width as well:

$$\langle VT \rangle^\gamma \sim \mathcal{N}(\mu_{\mathrm{VT}}^\gamma, \sigma_{\mathrm{VT}}^\gamma). \tag{B1}$$

Large Monte Carlo uncertainties in ⟨VT⟩ in some bins, due to the sparsity of simulated sources, can lead to inaccurate estimation of the astrophysical rate densities in said bins. To get around this problem, it is possible to marginalize our posterior distribution over the mentioned uncertainties in ⟨VT⟩ estimation (Farr 2019).

In order to carry out such a marginalization, we first need to estimate $\mu_{\mathrm{VT}}^\gamma$ and $\sigma_{\mathrm{VT}}^\gamma$. Following Farr (2019), we approximate these two quantities from the single realization of the samples used to compute the Monte Carlo sum, as in:

$$\mu_{\mathrm{VT}}^\gamma \approx \sum_{j \sim (\mathrm{det, draw})} \frac{\delta^{\gamma, \alpha(m_{1j}, m_{2j}, z_j)} T_{\mathrm{obs}}}{K_{\mathrm{draw}}} \\ \times \frac{\frac{dV_c}{dz}|_{z=z_j} \frac{1}{1+z_j} p(\vec{\lambda}_j | \mathrm{draw})}{p(m_{1j}, m_{2j}, z_j, \vec{\lambda}_j | \mathrm{draw}) m_{1j} m_{2j}}, \tag{B2}$$

$$(\sigma_{\mathrm{VT}}^\gamma)^2 \approx \sum_{j \sim (\mathrm{det, draw})} \frac{\delta^{\gamma, \alpha(m_{1j}, m_{2j}, z_j)} T_{\mathrm{obs}}^2}{K_{\mathrm{draw}}^2} \\ \times \left[ \frac{\frac{dV_c}{dz}|_{z=z_j} \frac{1}{1+z_j} p(\vec{\lambda}_j | \mathrm{draw})}{p(m_{1j}, m_{2j}, z_j, \vec{\lambda}_j | \mathrm{draw}) m_{1j} m_{2j}} \right]^2 - \frac{(\mu_{\mathrm{VT}}^\gamma)^2}{K_{\mathrm{draw}}}. \tag{B3}$$

With these estimates of $\mu_{\mathrm{VT}}^\gamma$ and $\sigma_{\mathrm{VT}}^\gamma$, it is possible to compute the posterior distribution of the rate densities and hyperparameters that have been marginalized over the Monte Carlo uncertainties in the ⟨VT⟩ estimation:

$$p(\vec{n}, \vec{\mu}, \sigma, \vec{\lambda} | \{w_d\}, \{\mu_{\mathrm{VT}}^\gamma\}, \{\sigma_{\mathrm{VT}}^\gamma\}) \\ = \int p(\vec{n}, \vec{\mu}, \sigma, \vec{\lambda} | \{w_d\}, \{\langle VT \rangle\}) \\ \times \left( \prod_\gamma d \langle VT \rangle^\gamma p(\langle VT \rangle^\gamma | \mu_{\mathrm{VT}}^\gamma, \sigma_{\mathrm{VT}}^\gamma) \right), \tag{B4}$$

where $p(\langle VT \rangle^\gamma | \mu_{\mathrm{VT}}^\gamma, \sigma_{\mathrm{VT}}^\gamma)$ is the normal distribution in Equation (B1). The integral in Equation (B4) can be evaluated analytically to obtain

$$\log p(\vec{n}, \vec{\mu}, \sigma, \vec{\lambda} | \{w_d\}, \{\mu_{\mathrm{VT}}^\gamma\}, \{\sigma_{\mathrm{VT}}^\gamma\}) \\ = \log p(\vec{\mu}, \sigma, \vec{\lambda}) - \frac{1}{2} \log |\Sigma(\sigma, \lambda)| \\ - \frac{1}{2} \sum_{\gamma, \gamma'} \{(\log n_\gamma - \mu_\gamma)(\log n_{\gamma'} - \mu_{\gamma'}) \Sigma_{\gamma\gamma'}^{-1}(\sigma, \lambda)\} \\ - \sum_\gamma n^\gamma \mu_{\mathrm{VT}}^\gamma + \frac{1}{2} \sum_\gamma (n^\gamma \sigma_{\mathrm{VT}}^\gamma)^2 + \sum_i \log \left( \sum_\gamma w_{d_i}^\gamma n^\gamma \right) + \mathrm{const.} \tag{B5}$$

It can be seen in Equation (B5) that the marginalized posterior is not normalizable, due to the $+(n^\gamma)^2$ term. This implies that the expectation values of the hyperparameters $\vec{\mu}, \sigma, \lambda$ with respect to the marginalized posterior become arbitrarily large. This is corrected for by imposing an additional constraint on $\vec{n}^\gamma$ during sampling, which is:

$$n^\gamma \mu_{\mathrm{VT}}^\gamma \leqslant 2 \left( \frac{\mu_{\mathrm{VT}}^\gamma}{\sigma_{\mathrm{VT}}^\gamma} \right)^2 \Rightarrow N_{\mathrm{det}}^\gamma \leqslant 2 N_{\mathrm{eff}}^\gamma. \tag{B6}$$

This condition is similar to the one derived in Farr (2019) for parametric population inference. Here, $N_{\mathrm{eff}}^\gamma$ is the effective number of independent samples of simulated events in the γth bin. Hence, the condition in Equation (B6) is implicative of the fact that the effective number of simulated events in a bin required for accurate VT estimation should be higher than the expected number of detectable events in said bin. It can be imposed by rejecting samples of $n^\gamma$ for which the condition is not satisfied.

Given the small number of observations analyzed for each study described in Section 3 and the correspondingly large number of simulated events used in VT estimation, this condition and the marginalized posterior in Equation (B5) were not implemented. On the other hand, it was verified in post-processing that for each sample of $n^\gamma$, the condition $N_{\mathrm{det}}^\gamma << N_{\mathrm{eff}}^\gamma$ was satisfied automatically. However, as summarized in this appendix, future studies that might attempt to use a sparse data set of simulated events in conjunction with a large number of observations can straightforwardly implement the condition in Equation (B6) as well as the marginalized posterior in Equation (B5) to prevent inaccurate VT and, hence, rate estimation.

On the other hand, Essick & Farr (2022) note that marginalizing over Monte Carlo uncertainties in the manner described so far might lead to biases in the inferred population, which in the case of parametric modeling can only be resolved in computationally expensive ways. Their argument is based on the fact that Monte Carlo uncertainties in the selection function are correlated between different points in the space of the population hyperparameters. The marginalization procedure described in Farr (2019), on which ours is based, assumes that the Monte Carlo uncertainties corresponding to different values of the population hyperparameters are uncorrelated, which leads to biases in the inferred distribution as compared to using the point estimates directly, without marginalization (Essick & Farr 2022). Hence, Essick & Farr (2022) argue that the resolution of this bias requires one to either use a much larger number of samples, as compared to the scenario wherein the point estimates are used directly, or to account for the aforementioned correlations by evaluating the point estimates on a multidimensional grid of population hyperparameters.

Given that both of these methods are expensive to implement for parametric models, they recommend the use of point estimates directly, instead of marginalization.

However, these concerns are not applicable in the context of our population model, due to the following reason. In our case, the "population hyperparameters" are the rate densities themselves. As described before, the mean and variance of the number of detectable events, as a function of the rate densities, are just sums of the rate densities raised to sum power, weighted by *precomputable* Monte Carlo integrals. Hence, the first concern raised by Essick & Farr (2022) is not applicable to our model, since the Monte Carlo sums over detectable simulations are only computed once and, unlike parametric models, not for every draw from the hyperposterior, leading to the net computational cost of our analyses remaining unchanged irrespective of how many detectable samples are used.

Similarly, accounting for correlations while marginalizing over the Monte Carlo uncertainties in our model can be implemented without any increase in computational cost. In the context of our model, correlations in the selection function between two points in the space of the population hyperparameters translate to correlations between the $\langle VT \rangle$ estimates of different bins. Hence, unlike parametric models, we need not compute Monte Carlo sums on a multidimensional grid of population hyperparameters for each draw from the hyperposterior. Instead we need only compute them once, that too only for all possible bin pairs.

For example, to account for these correlations, we need to replace Equation (B1) with a multivariate correlated Gaussian:

$$\{\langle VT \rangle\} \sim \mathcal{N}(\{\mu_{VT}\}, \Sigma_{VT}), \qquad (B7)$$

where $\Sigma_{VT}$ is the covariance matrix of the point estimates among the different bins. Following Essick & Farr (2022), we can estimate this covariance matrix from the Monte Carlo samples in the following way:

$$\Sigma_{VT}^{\gamma,\gamma'} = \frac{1}{K_{draw}} \left[ \frac{K_{det}}{(K_{det} - 1)K_{draw}} \sum_{j \sim (det,draw)} \right.$$
$$\times \left\{ \left( \delta^{\gamma,\alpha(m_{1j},m_{2j},z_j)} \frac{T_{obs} \frac{dV_c}{dz}|_{z=z_j} \frac{1}{1+z_j} p(\vec{\lambda}_j|draw)}{p(m_{1j}, m_{2j}, z_j, \vec{\lambda}_j|draw) m_{1j} m_{2j}} - \frac{K_{det}}{K_{draw}} \mu_{VT}^{\gamma} \right) \right.$$
$$\times \left. \left( \delta^{\gamma',\alpha(m_{1j},m_{2j},z_j)} \frac{T_{obs} \frac{dV_c}{dz}|_{z=z_j} \frac{1}{1+z_j} p(\vec{\lambda}_j|draw)}{p(m_{1j}, m_{2j}, z_j, \vec{\lambda}_j|draw) m_{1j} m_{2j}} - \frac{K_{det}}{K_{draw}} \mu_{VT}^{\gamma'} \right) \right\}$$
$$+ \left. \frac{K_{draw} - K_{det}}{K_{det}} \mu_{VT}^{\gamma} \mu_{VT}^{\gamma'} \right]. \qquad (B8)$$

Note that Equation (B8) is precomputable, given samples of simulated events. Once the covariances are estimated, one can proceed to carry out the marginalization in Equation (B4) with the correlated multivariate Gaussian of Equation (B7):

$$p(\vec{n}, \vec{\mu}, \sigma, \vec{\lambda}|\{w_d\}, \{\mu_{VT}^{\gamma}\}, \{\sigma_{VT}^{\gamma}\})$$
$$= \int p(\vec{n}, \vec{\mu}, \sigma, \vec{\lambda}|\{w_d\}, \{\langle VT \rangle\}) p(\{\langle VT \rangle\}|\{\mu_{VT}\}, \Sigma_{VT})$$
$$\times \left( \prod_{\gamma} d\langle VT \rangle^{\gamma} \right). \qquad (B9)$$

As before, the marginalization integral can be carried out analytically to obtain the following expression of the posterior:

$$\log p(\vec{n}, \vec{\mu}, \sigma, \vec{\lambda}|\{w_d\}, \{\mu_{VT}^{\gamma}\}, \Sigma_{VT}^{\gamma\gamma'})$$
$$= \log p(\vec{\mu}, \sigma, \vec{\lambda}) - \frac{1}{2} \log |\Sigma(\sigma, \lambda)| - \frac{1}{2} \sum_{\gamma, \gamma'} \{(\log n_{\gamma} - \mu_{\gamma})(\log n_{\gamma'} - \mu_{\gamma'}) \Sigma_{\gamma\gamma'}^{-1}(\sigma, \lambda)\}$$
$$- \sum_{\gamma} n^{\gamma} \mu_{VT}^{\gamma} + \frac{1}{2} \sum_{\gamma, \gamma'} (n^{\gamma} n^{\gamma'} \Sigma_{VT}^{\gamma\gamma'})$$
$$+ \sum_{i} \log \left( \sum_{\gamma} w_{d_i}^{\gamma} n^{\gamma} \right) + const. \qquad (B10)$$

Similar to the case of uncorrelated marginalization, the posterior in Equation (B10) is not normalizable, unless we impose the following condition

$$n^{\gamma} \mu_{VT}^{\gamma} \leqslant 2 \sum_{\gamma'} (\mu_{VT}^{\gamma} \mu_{VT}^{\gamma'} [\Sigma_{VT}^{-1}]^{\gamma\gamma'}) \qquad (B11)$$

in the form of a likelihood penalization. Note that Equation (B11) holds the same meaning as Equation (B6), with the only difference being that the number of effective samples of simulated events in each bin is now calculated while accounting for correlations with other bins. As before, we do not impose this condition during sampling in the form of a likelihood cut, due to the small number of observations and the correspondingly large number of detectable simulations. However, we do verify in post-processing that this condition is automatically satisfied for all the rate density samples.

To summarize, marginalization over the Monte Carlo uncertainties is implementable straightforwardly within our framework. The concerns raised by Essick & Farr (2022) are not applicable to our binned model, since, correlated or otherwise, the Monte Carlo sums required to evaluate our marginalized posterior are precomputable. Hence, unlike parametric models, we can correctly implement marginalization over the Monte Carlo uncertainties in the manner recommended by Essick & Farr (2022) while suffering no increase in computational cost.

## Appendix C
## Effects of Changing the Choice of Binning

In this appendix, we summarize the results referred to in Section 2.6 to demonstrate that different bin choices yield consistent results. Given the scalability of the current implementation with the dimensionality of the parameter space and the number of bins along each parameter dimension, we carry out this study in the context of the two-dimensional version of our model that was used by Abbott et al. (2023) to infer nonparametric constraints on the CBC mass spectrum. The three-dimensional models discussed in this work reduce to the mentioned two-dimensional one upon fixing the redshift evolution of the CBC population to a function that is uniform in comoving volume and completely uncorrelated with the masses (Mohite 2022; Abbott et al. 2023).

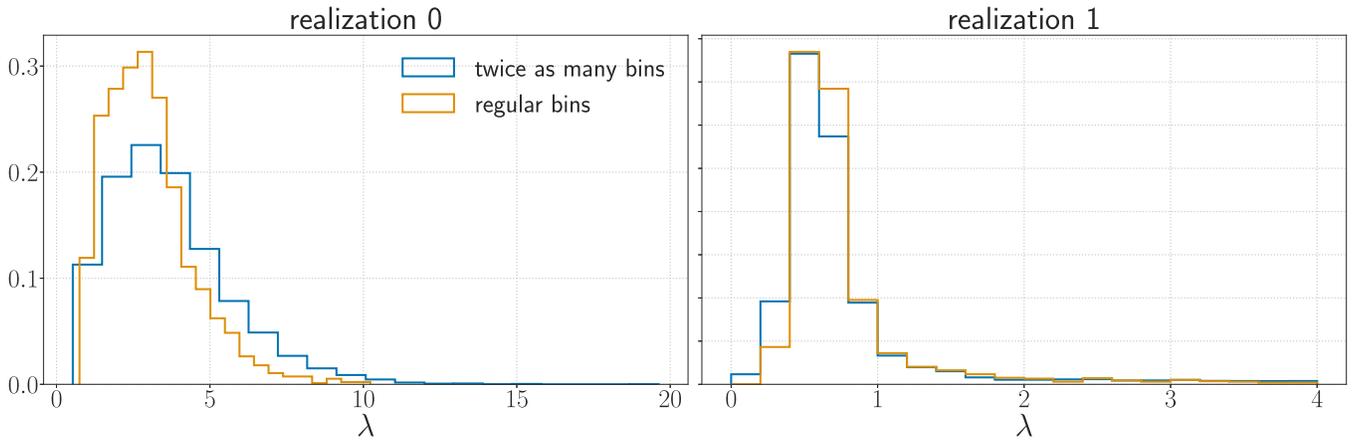

**Figure 7.** The effects of doubling the number of bins on the inferred length-scale posterior.

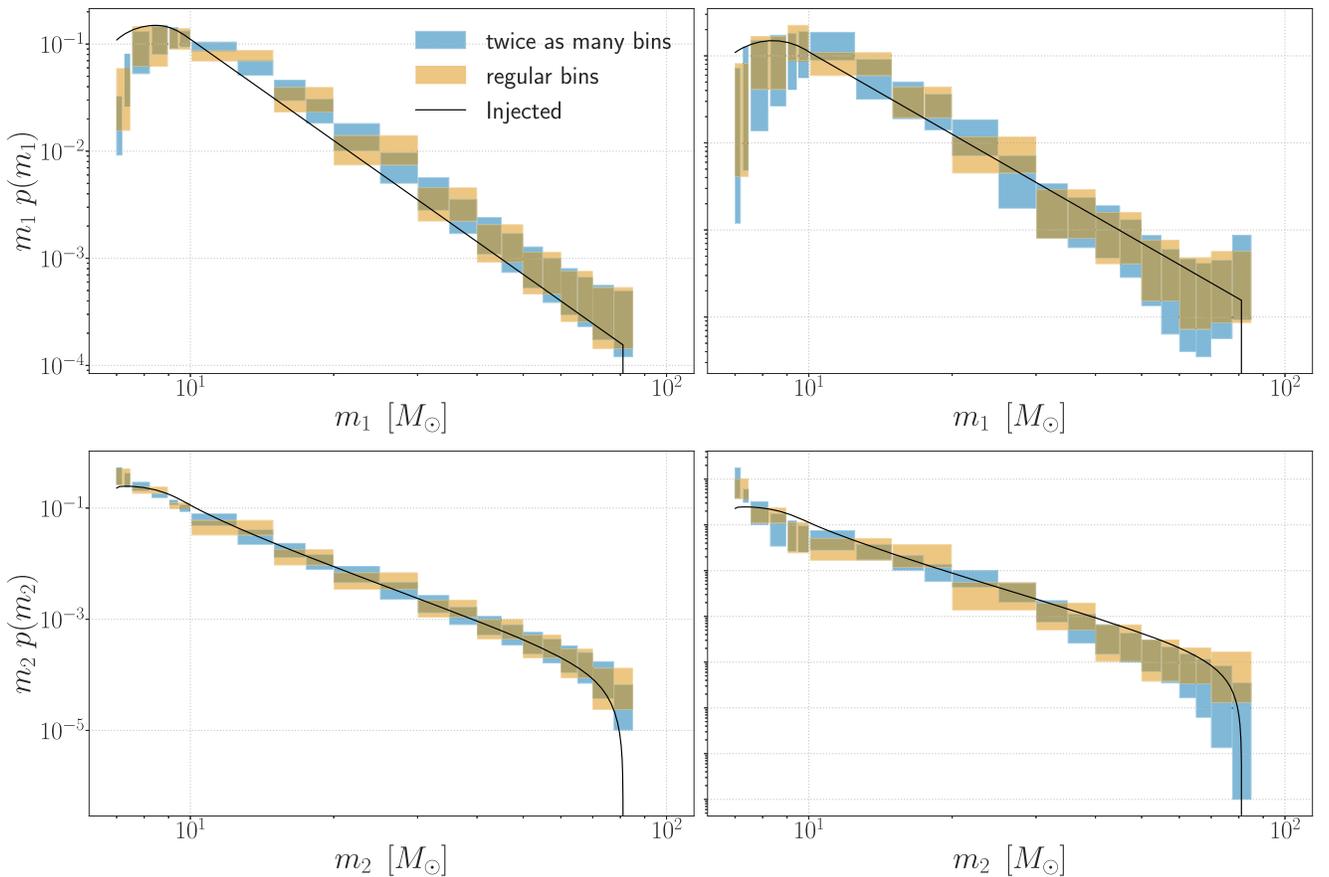

**Figure 8.** The effects of doubling the number of bins on the inferred population constraints.

For this study, we generate a simplistic population of simulated BBHs that is a smoothed, truncated power law in primary mass and mass ratio. The redshift evolution of the merger rate is chosen to be uniform, resulting in a redshift population that matches exactly with the assumptions of the abovementioned two-dimensional model. The functional form of the underlying population and the corresponding hyperparameters that characterize said function are specified in Equation (C1) and Table 3, respectively. We generate two realizations of this mock population, each comprising a set of 100 BBH events. We simulate the measurement uncertainty in the BBH parameters and detection sensitivity following the methodology described in Fishbach et al. (2018, 2020) and

Table 3
True Values for the Hyperparameters Characterizing the Underlying Population in Equation (C1)

| Hyperparameter | True Value |
| --- | --- |
| $\alpha$ | 3.14 |
| $\beta_q$ | 1.7 |
| $m_{\min}$ | $4.5 M_\odot$ |
| $m_{\max}$ | $55 M_\odot$ |

Farah et al. (2023), using the exact same procedure and power spectral densities summarized in Section 3.1.1.

$$\frac{dR}{dm_1 dm_2}(z) \propto m_1^{-\alpha} \left(\frac{m_2}{m_1}\right)^{\beta_q}. \tag{C1}$$

We analyze the aforementioned realizations of this simulated BBH population using the two-dimensional binned GP model corresponding to two different binning choices, with one containing twice as many bins as the other. The resulting inference is summarized in Figures 7 and 8.

It can be seen in Figure 7 that upon doubling the number of bins, the inferred posterior of the GP length scale remains unchanged, which is implicative of the GP correlating twice as many bins within the same interval of log-component masses. This leads to the inferred population constraints displayed in Figure 8, corresponding to the two different choices of binning, being fully consistent with each other, as well with the injected population up to measurement uncertainties. We thus conclude that increasing the bin resolution beyond what is needed to identify features in the underlying population leads to constraints that are consistent with those obtained from a lower-resolution inference, provided that the latter already captures all the features in the underlying population.

## ORCID iDs

Anarya Ray  https://orcid.org/0000-0002-7322-4748
Ignacio Magaña Hernandez  https://orcid.org/0000-0003-2362-0459
Siddharth Mohite  https://orcid.org/0000-0003-1356-7156
Jolien Creighton  https://orcid.org/0000-0003-3600-2406
Shasvath Kapadia  https://orcid.org/0000-0001-5318-1253